\documentclass[12pt]{elsarticle}

 \usepackage{graphics}

\usepackage{amssymb}

\biboptions{sort&compress}
\journal{Journal of Crystal Growth}

\begin{document}

\begin{frontmatter}



\title{Single crystal fiber growth of cerium doped strontium yttrate, SrY$_{2}$O$_{4}$:Ce$^{3+}$}


\author{J. Philippen}
\ead{jan.philippen@gmail.com}
\author{C. Guguschev}
\author{D. Klimm}

\address{Leibniz Institute for Crystal Growth, Max-Born-Str. 2, 12489 Berlin, Germany}

\begin{abstract}



First single crystal fibers of cerium doped strontium yttrate are fabricated using the laser-heated pedestal growth technique. Through thermodynamic equilibrium calculations and by high-temperature mass spectrometry suitable growth conditions can be determined. The atmosphere plays an important role during crystallization. It affects the composition shift, on the one hand, and the valence state of cerium, on the other hand. These properties can be explained by combining X-ray diffraction, elemental analysis, and optical spectroscopy. Crystallization in slightly reducing nitrogen atmosphere proves to be a reasonable choice, because evaporation is suppressed and trivalent cerium is stabilized. Strong green emission that depends on the oxygen fugacity during crystallization can be excited using UV light. Optical transmission of SrY$_{2}$O$_{4}$:Ce$^{3+}$ is measured for the first time.

\end{abstract}


\end{frontmatter}



\section{Introduction}
\label{intro}

This work is an approach to a new material, namely cerium doped strontium yttrate.
SrY$_{2}$O$_{4}$ is the only known stable interoxide within the system Y$_{2}$O$_{3}$-SrO~\cite{ropp2013}. The refractory oxide melts congruently at $\approx2215^{\circ}$\,C~\cite{tresvyatskii1971}. It was initially described in 1967~\cite{barry1967}. Its crystal structure was assigned to the calcium ferrate(III)-type, crystallizing in the orthorhombic space group \textit{Pnma}~\cite{muellerbuschbaum1968}. SrY$_{2}$O$_{4}$:Ce could be an appropriate UV to blue light converter~\cite{manivannan2003}. Moreover, the fabrication and characterization of single crystals are essential for new material research. 

Europium doped SrY$_{2}$O$_{4}$ has already been investigated as light phosphor~\cite{zhou2005, xu2001, park1999}.
Xu et al. described the optical properties of SrY$_{2}$O$_{4}$:Eu$^{3+}$~\cite{xu2001} confirming the assumption that trivalent europium ions are distributed on multiple sites~\cite{blasse1969}. Recently, additional REE doped SrY$_{2}$O$_{4}$ phosphors were investigated, including Tb$^{3+}$, Tm$^{3+}$, Dy$^{3+}$, and Sm$^{3+}$\cite{pavitra2012, zhang2013}.
Luminescence properties of cerium doped SrY$_{2}$O$_{4}$ were investigated by Blasse and Bril detecting a strong luminescence band with a maximum in the green wavelength region~\cite{blasse1967}. The authors also claim a remarkably large stokes shift.
 Manivannan at al. explain the emission spectrum of SrY$_{2}$O$_{4}$:Ce air fired samples with the Ce$^{3+}$ 5d-4f transition (blue band) and tentatively assign the  green emission to a Ce$^{4+}$-O$^{2-}$ charge transfer~\cite{manivannan2003}. 

The LHPG technique, based on the work of Burrus and Stone~\cite{burrus1975}, is suitable for the crystallization of SrY$_{2}$O$_{4}$. LHPG allows the application of almost arbitrary atmospheres, rapid crystallization rates, and effective distribution coefficients that are near unity.
In ideal steady state conditions a constant mass transfer from the feed into the melt zone and from the melt zone into the SCF is attained. In quasi-steady state conditions evaporating species generate a composition shift within the melt zone. However, a constant mass transfer into the fiber and therefore a constant composition will be possible. 

If trivalent cerium substitutes on the Y position of SrY$_{2}$O$_{4}$, no further charge compensation will be required. If Ce$^{3+}$ substitutes on the Sr position or Ce$^{4+}$ is incorporated, charge compensation will be required. A valence-coupled  diadochy can be formulated as follows according to~\cite{kroeger1956}: 
\begin{equation}
\mathrm{Ce}_{2}\mathrm{O}_{3}+3\mathrm{Sr}^{x}_{\mathrm{Sr}}=2\mathrm{Ce}^{\bullet}_{\mathrm{Sr}}+3\mathrm{SrO}+\mathrm{V}^{''}_{\mathrm{Sr}}.
\label{eq:kroeger_vink_1}
\end{equation}

Two trivalent cerium ions substitute two Sr$^{2+}$ ions within the SrY$_{2}$O$_{4}$ lattice. The valence is compensated by formation of strontium voids. The incorporation of tetravalent cerium on the trivalent Y position can be formulated likewise. Concerning the ion radii of the involved ions, tetravalent cerium is presumably incorporated on the Y position~\cite{shannon1976}. The Ce$^{3+}$ radius is about 101\,pm, if sixfold coordinated, and about 114\,pm, if eightfold coordinated. Thus, it is unclear, if Ce$^{3+}$ will be located on the Y side (90\,pm, sixfold) or on the Sr site (126\,pm, sixfold). 

\section{Experimental}
\label{exp}

\begin{table}[htb]
\small
\centering
\begin{tabular}{ll|ll}
\hline
Atmosphere                          	& Composition (mbar) 			&	Parameter          & LHPG 							\\
\hline
oxidizing I                        	 	& Ar (800), O$_2$ (200) 	& growth rate        & 0.1--2\,mm/min			\\
inert                              	 	& Ar (1000) 							& fiber diameter     & 0.5--3.0\,mm				\\
reducing I			                     	& N$_2$ (1000)  					& fiber length       & $\leq50$\,mm 			\\								
reducing II                        		& N$_2$ (950), H$_2$ (50) & aspect ratio $h/d_\mathrm{scf}$   & 1--3 \\ 
\hline
\end{tabular}
\caption{Composition of the atmospheres and other growth parameters for LHPG experiments. All gases had 5N (99.999\,\%) nominal purity. If a rest impurity is assumed as air, an oxygen partial pressure of $p_{\mathrm{O}_2}\approx2\times10^{-6}$\,bar can be estimated for all atmospheres not already having O$_2$ as a major component~\cite{klimm2009}.}
\label{tab:atmospheres}
\end{table}

The crystallization process occurred at $\approx2300^{\,\circ}$C. At these high temperatures thermodynamic equilibrium is reached quickly. Thermodynamic equilibrium calculations were performed to predict the valence of the dopant and the composition shift. FactSage\texttrademark~\cite{fact2013}, an integrated thermodynamic data bank system and Gibbs free energy minimization program, was used for these calculations. Several restrictions had to be acknowledged: (1) only a small volume of the system is at the estimated melt zone temperature; (2) within the melt zone there is a steep temperature gradient; (3) due to the high growth rates the system might not be in perfect equilibrium. Fugacities $f_{i}$ of evaporating species, the oxygen fugacity $f_{O_{2}}$ for each growth atmosphere, and the predominance diagram Sr-Y-Ce-O$_{2}$ were calculated dependent on the temperature. 

For fiber crystal  growth samples were prepared from SrCO$_3$, Y$_{2}$O$_3$ powders with 99.99\,\% purity. The dopant cerium was tetravalent CeO$_2$ (p.a. purity). Stoichiometric compositions of 0.02 CeO$_2$, 1.00 SrCO$_3$, 0.99 Y$_2$O$_3$ for cerium doped samples were calcinated at $1200^{\,\circ}$C for 10 hours. The products were multiply ground and sintered at $1200^{\,\circ}$C to ensure complete reaction to SrY$_{2}$O$_4$. Phase purity was checked by XRD. $\approx1.5$\,wt.$\%$  dissolved polyvinyl alcohol was added as organic binder. The wet powders were pressed isostatically at 2000\,bar in a die, sintered at $1200^{\,\circ}$C for 24\,h and cut into rectangular prisms. These prisms of polycrystalline SrY$_{2}$O$_4$ were used as feed rods and seed rods for LHPG. 

The LHPG furnace in this study is similar to the furnace described by Fejer et al.~\cite{fejer1984}. At the beginning of the LHPG process, the top of the feed rod was melted to a small droplet. With low translation rate the seed crystal was directed to the droplet and contacted the melt. After seeding, the feed rod and the seed rod have been moved upwards with similar translation rates until the fiber crystallized with a constant diameter. At the end of the process the seed rod was stopped and the fiber was pulled out of the melt. Table~\ref{tab:atmospheres} depicts the composition of the atmosphere and other growth parameters that were used for LHPG experiments. SCFs have been grown in flowing (up to 3\,Nl/min) and static atmosphere. The power of the laser as well as the translation rate of the pedestal were adjusted to assure an aspect ratio (h/r$_\mathrm{scf}$) between 1 and 3 (h: height of the melt zone, r$_\mathrm{cryst}$: diameter of the SCF). The melt zone was overheated approximately 100 to 150 K above the melting point of the material. The temperature was measured with a pyrometer. The edge length of the pedestal base ranged from 0.8 mm to 2.3 mm. 

Using an in-house built high-temperature mass spectrometer that is described in detail in~\cite{guguschev2012} the effective evaporation of SCFs was investigated. The fibers were heated up to $\approx2150^{\,\circ}$C in a tungsten crucible. The electron impact energy was set to 70\,eV and SEM voltage of 1300\,V was used for signal amplification. The temperature was adjusted manually and was measured with an optical pyrometer. 
For phase analysis an X-ray powder diffraction system (XRD 3003 TT, GE, USA) with Bragg-Brentano geometry was used. The X-Ray source was CuK$\alpha$ ($0.15406$\,nm).  
For elemental analysis via ICP-OES an IRIS Intrepid HR Duo (Thermo Elemental, USA) was used. 
For laser luminescence and lifetime measurements a nitrogen laser (MNL 200, LTB Lasertechnik Berlin, Germany) with an operating wavelength at 337.1\,nm and $\approx100$\,$\mu$J pulse energy was used. 
Transmission of plane parallel samples was measured between 190\,nm and 2500 \,nm (Lambda 19 UV/VIS/NIR, Perkin Elmer, USA). Samples were adjusted in front of a 1.0\,mm aperture. The slit width was 5\,nm and the measuring time set to 20\,nm/min. 

\section{Results and discussion}
\label{res}

\subsection{Thermodynamic considerations}


\begin{figure}
\centering
\includegraphics[width=1.0\textwidth]{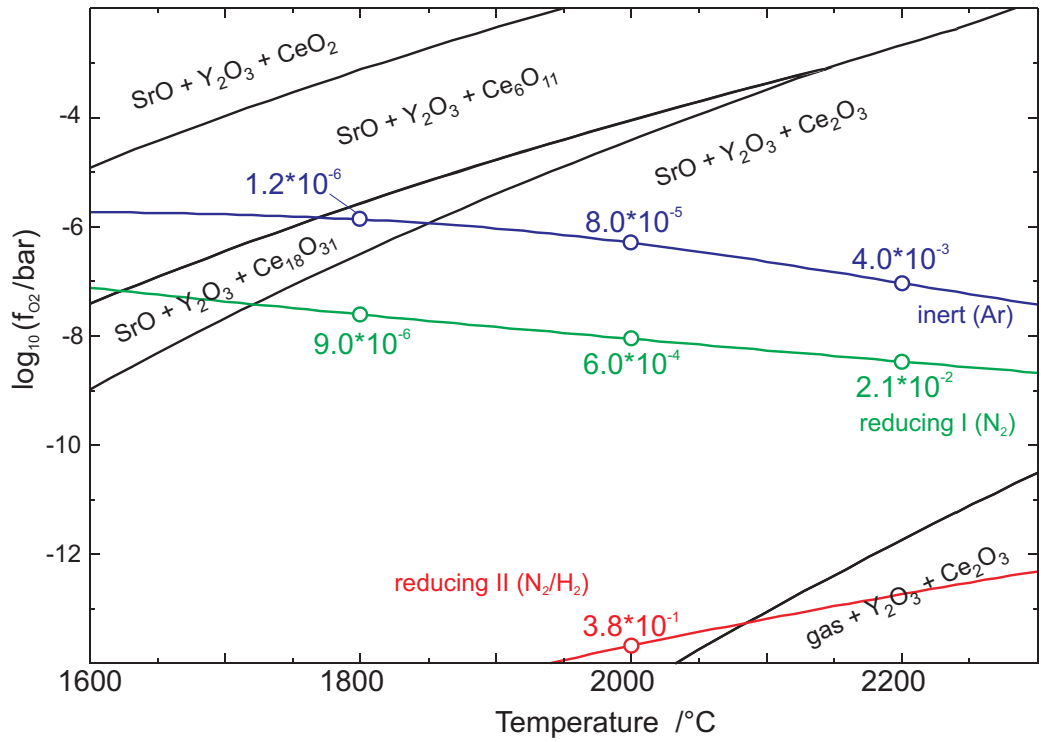} 
\caption{Predominance diagram Sr-Y-Ce-O$_{2}$ (black) showing the stability regions of trivalent and tetravalent cerium. The oxygen fugacities within the atmospheres that are used for crystal growth are superimposed. The numeric values represent the fugacities $f_{Sr}$ of the main evaporating species Sr.}
\label{fig:fact_atm1}
\end{figure}

\begin{figure}
\centering
\includegraphics[width=1.0\textwidth]{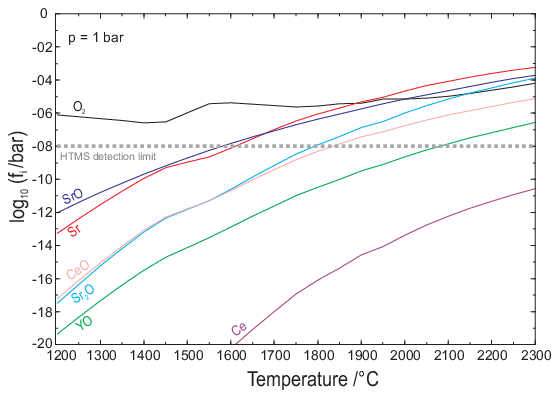} 
\caption{Fugacities $f_{i}$ of a SrY$_{1.98}$O$_{4}$:Ce$^{3+}_{0.02}$ composition in reducing atmosphere I (table~\ref{tab:atmospheres}) dependent on the temperature. The gray dashed line marks the corresponding HTMS detection limit.}
\label{fig:fact_sy2_1}
\end{figure}

The evaporation of components and the cerium valence are affected by the chemical nature of the atmosphere. The oxygen fugacity stabilizes either trivalent or tetravalent cerium. This can be explained by figure~\ref{fig:fact_atm1}. The predominance diagram Sr-Y-Ce-O$_{2}$ shows the regions of trivalent and tetravalent cerium dependent on the oxygen fugacity $f_{O_{2}}$ and the temperature. The elemental composition for the calculation and the composition that is used for crystal growth experiments are the same. For stabilizing trivalent cerium within a crystal matrix the oxygen fugacity of the growth atmosphere has to remain within the phase fields containing Ce$_{2}$O$_{3}$. This requirement is fulfilled for high temperatures in nitrogen containing atmospheres (reducing II, reducing I) and the inert argon atmosphere that are superimposed in figure~\ref{fig:fact_atm1}. For lower temperatures the fugacity of the inert atmosphere (blue) crosses the stabilization field tetravalent cerium. The nitrogen atmosphere (green) is slightly reducing compared to the inert atmosphere. The numeric values in figure~\ref{fig:fact_atm1} represent the fugacities $f_{Sr}$ of the main evaporating species Sr. Within the reducing atmosphere I $f_{Sr}\approx6.0*10^{-4}$\,bar at $2000^{\,\circ}$C, in comparison to $f_{Sr}\approx3.8*10^{-1}$\,bar for the reducing atmosphere II. Combining the oxygen fugacity and the $f_{Sr}$ values, the reducing atmosphere I seems to be appropriate for crystal growth. The low oxygen fugacity is stabilizing trivalent cerium within a large temperature range. Moreover, the evaporation of strontium, which depends on $f_{Sr}$, will be significant lower, if compared to the reducing atmosphere II (red). 

The fugacities of relevant evaporating species within the nitrogen atmosphere (reducing I), dependent on the temperature, are depicted in figure~\ref{fig:fact_sy2_1}. Despite minor deviations the sequence and ratios of $f_{i}$ do not alter with the temperature. At temperatures above $1600^{\,\circ}$C the strontium fugacity possesses the largest values, followed by SrO, Sr$_{2}$O, CeO and YO (decreasing in intensity). Strong evaporation of the alkaline earth metal and its oxides can be expected. The fugacity of CeO is almost two orders of magnitude smaller than Sr/SrO, but the initial Ce content within the fiber is also two orders of magnitude smaller. Therefore, a comparable deficit of Sr and Ce within the fibers can be expected. 

\begin{figure}[htb]
\centering
\includegraphics[width=1.0\textwidth]{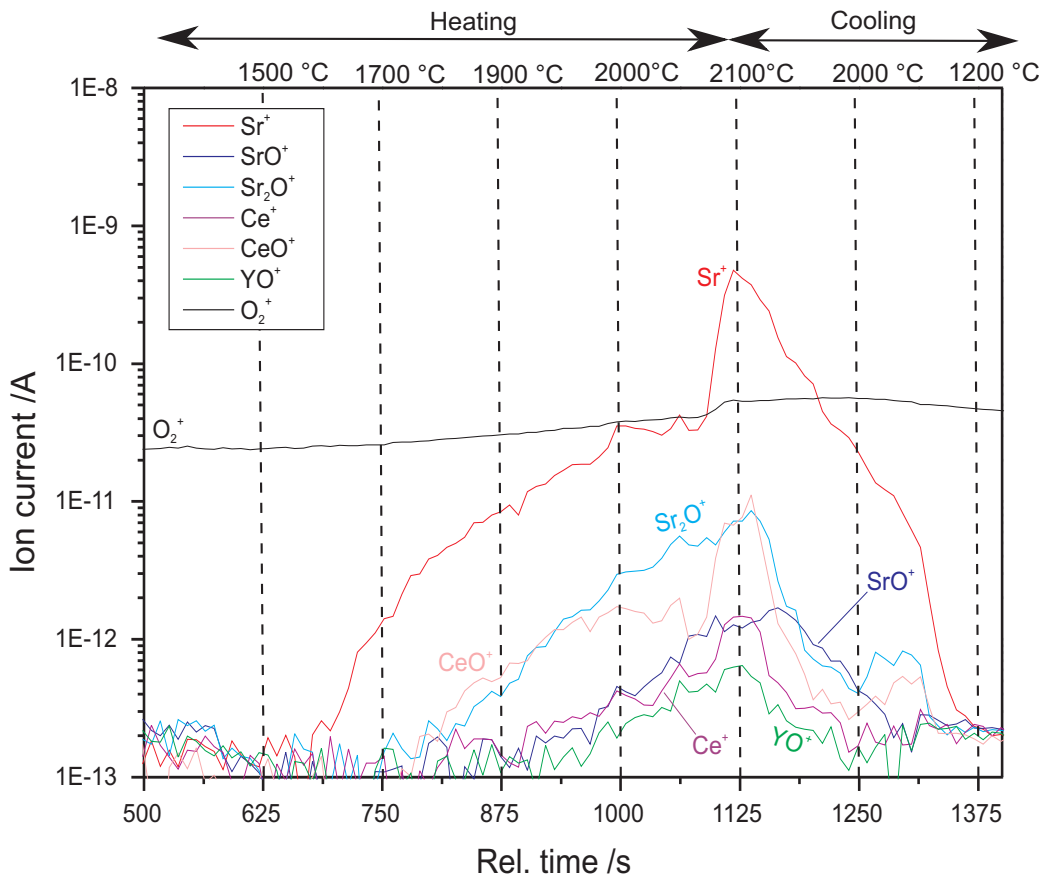} 
\caption{Monitored ion intensities of O$^{+}_{2}$ (black), Sr$^{+}$ (red), SrO$^{+}$ (blue), Sr$_{2}$O$^{+}$ (cyan), Ce$^{+}$ (purple), CeO$^{+}$ (rose), and YO$^{+}$ (green) during heating of a SrY$_2$O$_4$:Ce SCF up to $2150^{\,\circ}$C. The dashed lines show the temperature, which was adjusted manually.}
\label{fig:htms_sy2}
\end{figure}

To confirm these evaporating species high-temperature mass spectrometry was performed. Note that the HTMS measurement is kinetically oriented, whereas the fugacities are calculated for thermodynamical equilibrium conditions. Monitored ion intensities of O$^{+}_{2}$ (black), Sr$^{+}$ (red), SrO$^{+}$ (blue), Sr$_{2}$O$^{+}$ (cyan), Ce$^{+}$ (purple), CeO$^{+}$ (rose), and YO$^{+}$ (green) during heating of a SrY$_2$O$_4$:Ce SCF are depicted in figure~\ref{fig:htms_sy2}. The evaporating species are determined directly during heating of crystal fibers. Noticeable evaporation of Sr starts at a temperature of $\approx1600^{\,\circ}$C, which is far below the melting point of the material (figure~\ref{fig:htms_sy2}). The onsets of further detected evaporated species are Sr$^{+}$ at $\approx1600^{\,\circ}$C, Sr$_{2}$O$^{+}$, SrO$^{+}$ and CeO$^{+}$ at $\approx1750^{\,\circ}$C, Ce$^{+}$ at $\approx1800^{\,\circ}$C, and YO$^{+}$ at $\approx1950^{\,\circ}$C. These onsets can be compared with the fugacities $f_{i}$ (in figure~\ref{fig:fact_sy2_1}) for each evaporating species at the same temperature. For each temperature onset of the ion intensity the corresponding fugacity value can be marked at $\approx10^{-8}$\,bar. This fugacity value is marked as gray dashed line in figure~\ref{fig:fact_sy2_1} (HTMS detection limit). Note that this limit is valid for a sample weight of $\approx50$\,mg. 

HTMS measurements could verify the appearance of all calculated species experimentally. Moreover, the onsets of the detected ion intensities match the sequence of the calculated fugacities properly. A minor difference in the sequence might arise from different ion cross sections or fragmentation resulting in stronger Sr$_{2}$O$^{+}$, CeO$^{+}$, and Ce$^{+}$ signals.
For a quantitative analysis calibration runs are necessary and the ion cross sections and fragmentation of the monitored species have to be considered. Fugacities could be determined experimentally by measuring the partial pressures of the gaseous species by Knudsen effusion mass spectrometry (KEMS) under thermodynamical equilibrium conditions. 

For crystal growth experiments, the following conditions can be summarized: Regarding the evaporation, a slightly oxidizing atmosphere (oxidizing I) seems to be appropriate. But for stabilizing trivalent cerium a reducing atmosphere is required. If a valence-coupled diadochy for the cerium doping is performed by the generation of Sr vacancies (equation~\ref{eq:kroeger_vink_1}), the Sr evaporation will promote the Ce$^{3+}$ incorporation. This is also suggested for the CaSc$_{2}$O$_{4}$:Ce phosphor and SCF~\cite{shimomura2007, philippen2013}. The influence of the growth atmosphere on the crystal properties will be presented in the following. 

\subsection{Crystal growth}


\begin{figure}[htb]
\centering
\includegraphics[width=1.0\textwidth]{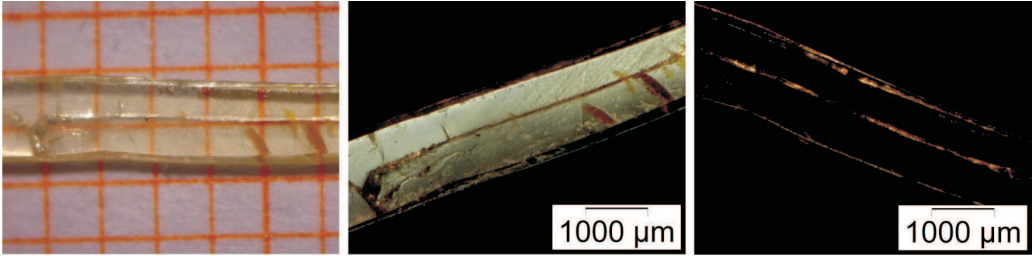} 
\caption{A yellow-green SrY$_{2}$O$_{4}$:Ce SCF (left, upon mm scaled paper). The plane parallel polished fiber reveals homogeneous extinction between cross-polarized light (mid, right).} 
\label{fig:diss1_scf_sy2_1}
\end{figure}

\begin{table}[htb]
\small
\centering
\begin{tabular}{ll}
\hline
Atmosphere 								&	reducing I										\\ 
$\nu_{\mathrm{seed}}$ 							&	1\,mm/min											\\
$\nu_{\mathrm{feed}}$ 							&	1\,mm/min											\\
$r_{\mathrm{feed}}$ 								& $\approx1.25$\,mm							\\
$r_{\mathrm{scf}}$									& $\approx1.00$\,mm							\\
aspect ratio 												&	$\approx1.5$									\\ 
$T_{\mathrm{max}}$ 									&	$\approx2300^{\,\circ}$C			\\ 
\hline 
\end{tabular}
\caption{The optimized LHPG growth conditions for SrY$_{2}$O$_{4}$:Ce$^{3+}$ SCFs.}
\label{tab:opti_conditions}
\end{table}

\begin{figure}[htb]
\centering
\includegraphics[width=1.0\textwidth]{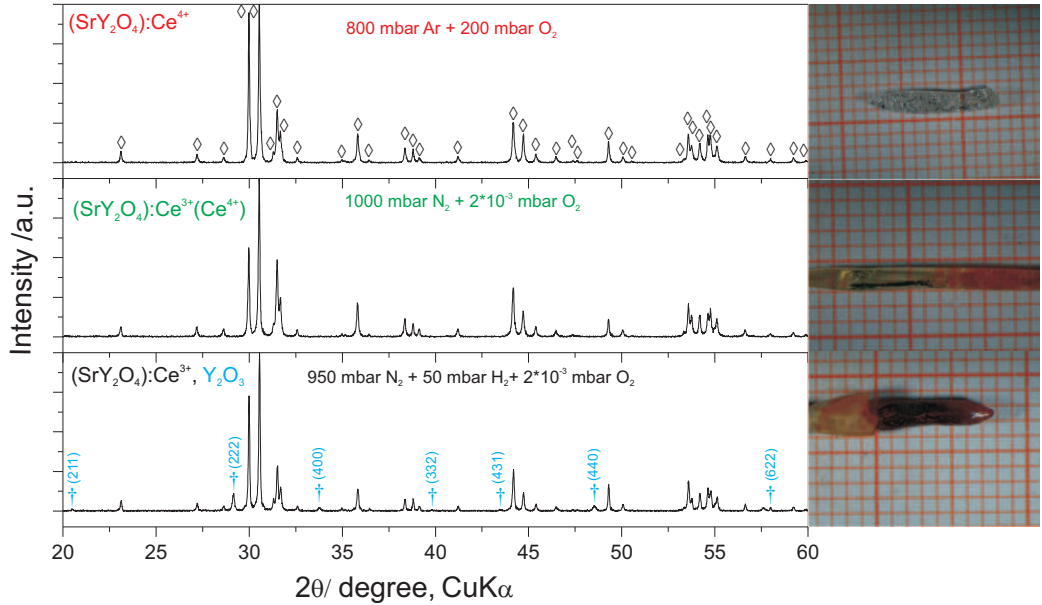} 
\caption{XRD patterns of three crystal fibers that has been crystallized in different atmospheres: oxidizing I (red, top), reducing I (green, mid), reducing II (black, bottom). A representative two theta range has been selected.} 
\label{fig:sy2_atm_pic2}
\end{figure}

Strontium yttrate was grown with the crystal growth conditions given in table~\ref{tab:atmospheres}. A yellow-green colored SCF, with a visible crack along the growth direction, shows homogeneous extinction between cross-polarized light (figure~\ref{fig:diss1_scf_sy2_1}).  The crystallized fibers did not show black color that could be related to oxygen vacancy formation.

Figure~\ref{fig:sy2_atm_pic2} shows the XRD patterns of strontium yttrate fibers crystallized in different atmospheres. If grown in oxidizing atmosphere (top), fibers will be fully transparent. The XRD patterns do not indicate a CeO$_{2}$ phase impurity. Moreover, all reflections can be indexed with PDF No. 01-74-264~\cite{powder2011}. If crystallized in nitrogen atmosphere (mid), the fibers will be green-yellow colored and homogeneous. Crack-free SCFs could be crystallized in this atmosphere. If hydrogen is added, fibers will be heterogeneous with phase impurities of Y$_{2}$O$_{3}$ (bottom). The best quality is attained in slightly reducing nitrogen atmosphere. The optimized growth parameters are given in table~\ref{tab:opti_conditions}. In oxidizing atmosphere, fibers tend to crack. This is probably caused by melt oscillations during crystallization, as it is observed for calcium scandate growth~\cite{philippen2013}. These oscillations only occur in oxidizing atmosphere. Strong evaporation causes a composition shift in reducing atmospheres containing hydrogen. Thus, an additional phase Y$_{2}$O$_{3}$ is crystallized. Dendritic formation of SrO needles at the growth interface will interfere the crystallization process, if hydrogen is present, as it is observed for calcium scandate crystallization, likewise~\cite{philippen2013}.

\begin{table}[htb]
\small
\centering
\begin{tabular}{llllr}
\hline
\multicolumn{1}{c}{Sample}            & \multicolumn{3}{c}{Rel. deviation (\%)} & \multicolumn{1}{r}{Sample number}  \\ 
\hline 
                                      &  Sr   &   Y             		 &  Ce       &	 \\
\hline
Feed rod                              & -0.3	& +0.2			        & +1.21	   		& SY2-P1\\
\hline
SCF, oxidizing I, begin 		       		& -0.5  & +0.8              & -54.5     	& SY2-6 \\
SCF, oxidizing I, end    							& -2.5  & +1.7              & -40.5     	& SY2-6 \\
FZ, oxidizing I                				& +2.0 	& -1.3              & +26.4     	& SY2-6 \\
\hline
SCF, reducing I, begin 		       	  	& -1.7  & +1.2              & -39.0     	& SY2-5 \\
SCF, reducing I, end    							& -1.6  & +1.1              & -26.2     	& SY2-5 \\
FZ, reducing I                				& -6.0 	& +2.6              & +48.2     	& SY2-5 \\
\hline
CF, reducing II, begin 		       		 	& -68.8 & +34.6            	& 14.8     		& SY2-3 \\
CF, reducing II, end    							& -37.8 & +19.2             & -14.7     	& SY2-3 \\
FZ, reducing II                				& -2.9	& +2.0              & -53.5     	& SY2-3 \\
\hline 
\end{tabular}
\caption{Relative deviation (from the weighed portion) of the elements within the feed rod, crystal fibers (SCF, CF), and corresponding frozen melt zones (FZ). The first column marks the growth atmosphere and the fiber position (begin/end). The elemental analysis was performed using ICP-OES.}
\label{tab:sy2_deviation}
\end{table}

The growth parameters affect the dopant distribution and the content of the major elements. The composition shifts of an air-sintered feed rod, crystal fibers, and their corresponding frozen melt zones are presented in table~\ref{tab:sy2_deviation}. Some strontium evaporated already during the preparation of the feed rod. This loss can be minimized by reduced sintering temperature and duration. During the crystallization strontium (and its oxides) and cerium evaporated. Condensed evaporated species could be collected and determined as mainly SrO.  While the deviations in oxidizing conditions (oxidizing I ) and nitrogen atmosphere (reducing I) are similar, a hydrogen containing atmosphere strongly interferes the composition. Large amounts of Ce and Sr evaporated and an SCF could not be fabricated. These results are in accordance with the HTMS measurements and the calculated fugacities. The deviations are minimized in nitrogen atmosphere (reducing I). Moreover, the deviations of the main elements Sr and Y are constant. The cerium content within the fiber increases from the begin ($-39.0$\,\%) to the end ($-26.2$\,\%). Thus, in reducing atmosphere I, a quasi-steady state is attained with respect to the main elements. The cerium content is not only affected by the evaporation, but also by segregation.  

For LHPG setups the effective distribution coefficient $k_\mathrm{eff}$ can be estimated by~\cite{wilke1988}:  
\begin{equation}
k_\mathrm{eff} \approx \frac{C_\mathrm{ini}}{C_0} \approx \frac{C_\mathrm{end}}{C_\mathrm{fz}}
\label{eq:lhpg_pfann}
\end{equation}
with $C_\mathrm{ini}$ and $C_\mathrm{end}$ as dopant concentration at the beginning and the end of the fiber, $C_0$ as dopant concentration within the feed rod, and $C_\mathrm{fz}$ as dopant concentration in the frozen melt zone. Based on this equation we can estimate $k_\mathrm{eff}\approx0.5$ for crystallization in oxidizing atmosphere and $k_\mathrm{eff}\approx0.6$ for crystallization in nitrogen atmosphere. Note that equation~\ref{eq:lhpg_pfann} will be imprecise, if strong evaporation occurs. In this case, the steady state condition ($C_\mathrm{end} = {C_0}$) is never reached~\cite{philippen2013}.
The average effective distribution coefficient is slightly higher than for the cerium distribution in calcium scandate~\cite{philippen2013}. This difference might arise from different doping mechanism. In calcium scandate cerium incorporates on the Ca position. In SrY$_{2}$O$_{4}$ cerium can be incorporated on the Y position without charge compensation.

\begin{figure}[htb]
\centering
\includegraphics[width=1.0\textwidth]{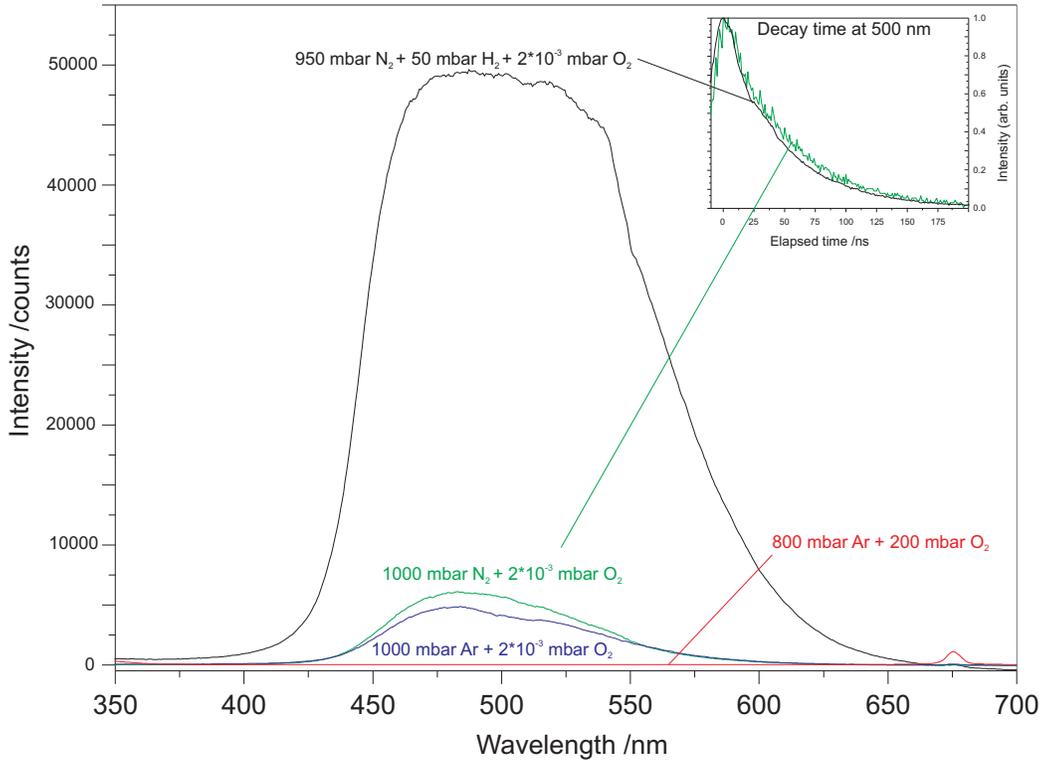} 
\caption{Laser fluorescence spectra and radiative lifetime of crystal fibers that were crystallized in different atmospheres: reducing II (black), reducing I (green), inert (blue), oxidizing I (red). The fluorescence excitation was at 337.1\,nm. The radiative lifetime was determined at the luminescence maximum.} 
\label{fig:sy2_atm1}
\end{figure}

The laser fluorescence emission and radiative lifetimes of crystal fibers that were grown in different atmospheres are depicted in figure~\ref{fig:sy2_atm1}. If fibers are crystallized in strongly reducing atmosphere (black curve), a broad emission band will be excited with a peak wavelength at 489\,nm. The location and structure of this band is similar to the luminescence from powder samples~\cite{blasse1967}. The radiative lifetime does not depend on the crystallization atmosphere and shows a single exponential decay with~$\tau=45\pm5$\,ns. The fluorescence curves and their radiative lifetime are in accordance with the reported blue emission band, caused by the 5d-4f Ce$^{3+}$ transition~\cite{manivannan2003}. We could not find evidence for any kind of Ce$^{4+}$-O$^{2-}$ charge transfer luminescence, as reported by the same authors~\cite{manivannan2003}. If fibers are crystallized in oxidizing atmosphere that stabilizes tetravalent cerium, luminescence will be quenched (red curve). In this case, all cerium is tetravalent. Nitrogen atmosphere during crystallization results in a luminescence (green curve) that led to considerable amount of trivalent cerium. However, a certain amount of tetravalent cerium has to be considered. The luminescence intensity seems to be related to the Ce$^{3+}$/Ce$^{4+}$ ratio that strongly depends on the oxygen fugacity, as reported in~\cite{philippen2013}. A proper accordance between the intensity of the luminescence and the Ce$^{3+}$/Ce$^{4+}$ ratio that can be assumed from the predominance diagram in figure~\ref{fig:fact_atm1} can be emphasized. 


\begin{figure}[htb]
\centering
\includegraphics[width=1.0\textwidth]{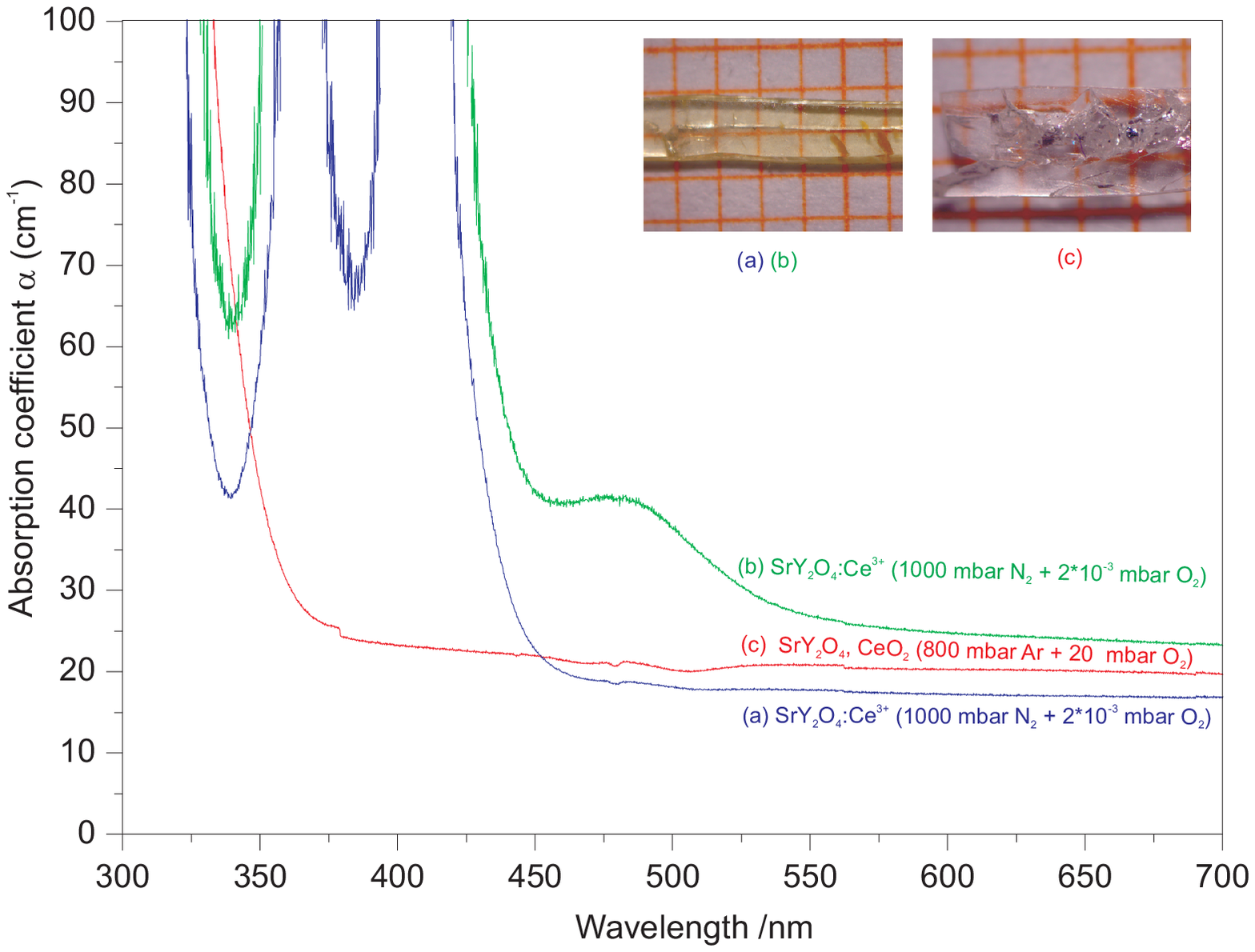} 
\caption{Absorption of SrY$_{2}$O$_{4}$:Ce crystals that has been crystallized in reducing atmosphere I (a,b) and in oxidizing atmosphere I (c). }
\label{fig:diss1_cs2_sy2_abs_1}
\end{figure}

The absorption coefficient can be calculated from transmission measurements. Absorption of SrY$_{2}$O$_{4}$:Ce crystal fibers is shown in figure~\ref{fig:diss1_cs2_sy2_abs_1}. The SCF that has been crystallized in nitrogen atmosphere reveals several absorption bands. Two bands, located at $\approx360\,$nm and $\approx400\,$nm, can be detected in curve (a), which was measured throughout the yellow-colored fiber at the begin of the crystallization process. Curve (b) was taken throughout an orange-colored part of the same crystal fiber showing an additional absorption band located at $\approx475\,$nm. The additional absorption band might be related to incorporation of cerium on an additional crystallographic site.
More precisely, cerium would be incorporated on the two Y sites in the host lattice. If the total cerium concentration exceeds a certain value, it could be also incorporated on the Sr site, resulting in absorption at larger wavelengths. This assumption would be in accordance with the europium doping mechanism for strontium yttrate~\cite{xu2001}.
The fiber including tetravalent cerium (red curve) reveals neither of the above mentioned absorption bands. The absorption can be compared to pure calcium scandate~\cite{fechner2011} and pure ceria, CeO$_{2}$~\cite{chevire2006}. It is yet unclear, if tetravalent cerium is incorporated into strontium yttrate. However, an additional CeO$_{2}$ phase could not be detected using XRD (figure~\ref{fig:sy2_atm_pic2}). All three absorption bands can be tentatively assigned to trivalent cerium or crystal structure defects that are caused by cerium doping.
Because of the high overlap of absorption and emission cerium doped strontium yttrium oxide does not seem to be an appropriate material for laser (self-absorption). Note that the spectroscopic measurements are only a first approach to investigate the precise doping process. 

\section{Conclusion and outlook}
\label{conc}

An approach to the investigation of a new material, SrY$_{2}$O$_{4}$:Ce$^{3+}$, is given. Using optimized growth parameters single crystal fibers, doped with trivalent cerium, can be fabricated. Through thermodynamic calculations, supported by HTMS and elemental analysis, suitable conditions for LHPG growth could be determined. The crystallization atmosphere turned out to be a critical growth parameter as it affects the composition shift (through evaporation) and the cerium valence (through oxygen fugacity). By crystallization in nitrogen atmosphere the evaporation of species is suppressed and trivalent cerium is stabilized. 
The amount of trivalent cerium can be estimated by laser luminescence intensity and absorption measurements.

A certain fraction of tetravalent cerium has to be taken into account. The accurate Ce$^{3+}$/Ce$^{4+}$ ratio, dependent on the oxygen fugacity, is yet unknown. For a precise analysis bulk crystals with higher quality are required. Thereby, Ce$^{3+}$/Ce$^{4+}$ could be analyzed using X-ray absorption spectroscopy~\cite{melcher2004}. Bulk crystal growth using Czochralski technique, based on this study, could be a suitable method for high quality crystals and further analysis.



\section*{Acknowledgments}

We would like to acknowledge the support of R. Bertram and A. Kwasniewski from the Leibniz Institute for Crystal Growth.



\bibliographystyle{model1-num-names}








\end{document}